\documentclass[usenatbib]{mnras}
\usepackage{graphicx, amsmath, amssymb}
\usepackage{url}
\usepackage{amsmath}
\usepackage[T1]{fontenc}
\usepackage{ae,aecompl}
\usepackage{color}

\newcommand{\feh}{{\rm [Fe/H]}}
\newcommand{\gyr}{{\, \rm Gyr}}
\newcommand{\kpc}{{\, \rm kpc}}
\newcommand{\pcd}{{\, \rm pc^{-3}}}

\newcommand{\rh}{r_{\rm h}}
\newcommand{\rc}{r_{\rm c}}

\newcommand{\rj}{r_{\rm J}}

\newcommand{\msun}{\, {\rm M}_{\sun}}
\newcommand{\lsun}{\, {\rm L}_{\sun}}
\newcommand{\trh}{\tau_{\rm rh}}
\newcommand{\trhz}{\tau_{\rm {rh,0}}}
\def\camcos{Commun. Appl. Math. Comput. Sci.}

\title[A stellar-mass BH population in NGC 6101?]{A stellar-mass black hole population in the globular cluster NGC 6101?}
\author[M. Peuten]{M. Peuten\thanks{E-mail:
m.peuten@surrey.ac.uk }, A. Zocchi, M. Gieles, A. Gualandris, V. H\'enault-Brunet\\
 Department of Physics, University of Surrey, Guildford  GU2 7XH, UK}

\begin{document}

\date{}

\pagerange{\pageref{firstpage}--\pageref{lastpage}} \pubyear{2015}

\maketitle

\label{firstpage}

\begin{abstract}
Dalessandro et al. observed a similar distribution for blue straggler stars and main-sequence turn-off stars in the Galactic globular cluster NGC 6101, and interpreted this feature as an indication that this cluster is not mass-segregated. Using direct $N$-body simulations, we find that a significant amount of mass segregation is expected for a cluster with the mass, radius and age of NGC 6101. Therefore, the absence of mass segregation cannot be explained by the argument that the cluster is not yet dynamically evolved. By varying the retention fraction of stellar-mass black holes, we show that segregation is not observable in clusters with a high black hole retention fraction (>50\% after supernova kicks and >50\% after dynamical evolution). Yet all model clusters have the same amount of mass segregation in terms of the decline of the mean mass of stars and remnants with distance to the centre. We also discuss how kinematics can be used to further constrain the presence of a stellar-mass black hole population and distinguish it from the effect of an intermediate-mass black hole. Our results imply that the kick velocities of black holes are lower than those of neutron stars. The large retention fraction during its dynamical evolution can be explained if NGC 6101 formed with a large initial radius in a Milky Way satellite. 
\end{abstract}


\begin{table*}
\caption{Properties of the globular cluster NGC 6101. The first line lists the current properties of the cluster, the second line the initial properties as determined with \textsc{emacss} and the third line the scaled initial properties that we use as initial conditions to set up our simulation. Columns from 2 to 10 list, respectively, the total $V$-band luminosity $L_{\rm V}$ of the cluster in solar units, the total mass of the cluster $M_{\rm cl}$ in solar masses, the number of stars $N$, the average mass of stars in solar masses, the half-mass radius $\rh$ in pc, the effective radius $R_{\rm eff}$ in pc, the Galactocentric distance $R_{\rm G}$ in kpc, the metallicity $\feh$ in solar units and the half-mass relaxation time $\trh$ in Gyr. The references for the values listed in the first line are indicated in the notes.}
\label{tab:initialconditions}
\begin{tabular}{cccccccccc}
\hline 
                            & $L_{\rm V}$               & $M_{\rm cl}$        & $N$                 & $\bar{m}$ & $\rh$            & $R_{\rm eff}$   & $R_{\rm G}$   & $\feh$                  & $\trh$ \\
\hline 
Current properties          & $5.7 \times 10^{4 \star}$ & $1.1 \times 10^{5}$ & $2.7 \times 10^{5}$ & $0.40$    & $12.3^{\dagger}$ & $9.2^{\dagger}$ & $11.2^{\star}$ & $-1.98 {\,}^{\ddagger}$ & $5.4-6.3 {\,}^{\dagger}$ \\
Initial unscaled properties & $-$                    & $2.0 \times 10^{5}$ & $3.1 \times 10^{5}$ & $0.64$    & $5.8$            & $4.4$           & $11.2$           & $-1.98$                 & $2.8$ \\
Initial scaled properties   & $-$                    & $6.4 \times 10^{4}$ & $1.0 \times 10^{5}$ & $0.64$    & $7.6$            & $5.7$           & $30.0$           & $-1.98$                 & $2.8$ \\
\hline 
\end{tabular}

References: ${\,}^{\star}$ \cite{1996AJ....112.1487H}, ${\,}^{\dagger}$ \cite{Dalessandro2015} and 
${\,}^{\ddagger}$ \cite{2009A&A...508..695C}.  
\end{table*} 

\section{Introduction}
\label{Intro}

Globular clusters (GCs) are old stellar systems ($\sim 10 - 13 \gyr$) with masses ($\sim$ few $10^5 \msun$) and densities ($\sim$ few $1000 \msun \pcd$) resulting in two-body relaxation time-scales shorter than their age. In two-body encounters, the lighter stars generally gain velocity while the heavier stars lose velocity. After many subsequent encounters, the low-mass stars gain velocity with respect to the high-mass stars, which in turn means that in GCs, light stars are found further away from the centre than high-mass stars \citep{1995ApJ...452L..33K}. This effect is generally referred to as mass segregation. Because GCs are older than their respective half-mass relaxation time ($\trh$) \citep{1961AnAp...24..369H,2011MNRAS.413.2509G}, we expect the stars and remnants of different masses to have different distributions in phase space. This effect has been confirmed observationally \citep{1995ApJ...452L..33K,2014MNRAS.437.1918S}. 

There are different ways to study mass segregation in GCs. The one we will mostly refer to in this study is the use of cumulative radial distributions of stars with different masses: in a mass-segregated cluster, we expect the stars with high mass, such as blue straggler stars (BSSs) to be more centrally concentrated than main-sequence turn-off stars (MSTO). If the cluster is not mass-segregated, then the cumulative radial distributions are the same. \cite{2008ApJ...686..303G} showed that the presence of an intermediate-mass black hole (IMBH) reduces the amount of mass segregation among observable stars, and they suggest that this could be used as an observable indication of the presence of an IMBH.

\cite{Dalessandro2015} (hereafter D15) recently studied different properties of the GC NGC 6101, such as the radial distribution of the BSSs, the radial variation of the binary fraction and the radial variation of the luminosity and mass function (MF). From their analyses, they conclude that this cluster is not mass-segregated. They also found a large core radius relative to the half-light radius (effective radius) for the GC ($R_{\rm c} / R_{\rm eff} \approx 0.4$).  

NGC 6101 is a metal-poor cluster with $\feh = -1.98$ \citep{2009A&A...508..695C} located at a distance of $14.6$ kpc (D15) from the Sun and $11.2$ kpc \citep{1996AJ....112.1487H} from the Galactic Centre. When fitting a \cite{1966AJ.....71...64K} model to the observed number density profile, D15 obtained a concentration $c = \log(r_{\rm t}/r_{\rm c}) = 1.3$ and a projected effective radius of $R_{\rm eff} = 128.2$ arcsec. These values are larger than the values listed in the \cite{1996AJ....112.1487H} catalogue and those given by \cite{2005ApJS..161..304M}. D15 attribute the larger radii to their improved method of background subtraction. D15 estimate that NGC 6101 has an half-mass relaxation time-scale of $\trh \sim 5.4 - 6.3$ Gyr. 

The value of the initial half-mass relaxation time ($\trhz$) for NGC 6101 should be smaller than the one we measure today for this cluster, because in roughly the first half of the evolution of tidally limited GCs, the half-mass relaxation time increases due to stellar mass-loss and two-body relaxation-driven expansion \citep{2010MNRAS.408L..16G}. In Section~\ref{sec:Nbody}, we estimate $\trhz$ to be $\sim 2.8 \gyr$. \cite{2008ApJ...686..303G} found that a cluster needs to be $\sim 5 \trhz$ old to appear fully mass-segregated. Given the estimated age of $13 \gyr$ \citep{2010ApJ...708..698D}, we expect NGC 6101 to show signs of mass segregation.

The objective of this study is to understand this contradiction: on one hand, the cluster appears to be not mass-segregated, on the other hand mass segregation is expected based on the age and estimated $\trhz$. It has been shown \citep{2004MNRAS.355..504M,2004ApJ...608L..25M,2013A&A...558A.117L} that a population of heavy remnants can result in a large core radius ($\rc$) over half-mass radius ($\rh$), as observed for NGC 6101. Because black hole (BH) candidates have recently been observed in several GCs \citep{2012Natur.490...71S,2013ApJ...777...69C}, we investigate the effect of a population of remnants on the apparent mass segregation for the case of NGC 6101. To do this, we use a set of $N$-body simulations with different retention fractions of BHs, and we compare them to the observations. We also use dynamical equilibrium models to formulate predictions on other observable quantities. 

This paper is organized as follows: in Section \ref{sec:Nbody}, we present the $N$-body models used in this analysis. In Section \ref{sec:Results}, we discuss the results of the analysis of our $N$-body models and we show the effects produced by a population of stellar-mass BHs on the observations. In Section \ref{sec:discussion}, we propose a method to observationally distinguish the scenario we introduce here from other possible explanations, by looking at the kinematics of the cluster. In Section \ref{sec:Conclusion}, we discuss our results in the context of other scenarios and present our conclusions.

\begin{table*}
\caption{Initial and final properties of the three $N$-body models, as indicated in the first column. We list the values of the number 
of bound stars $N$, the total mass of bound stars $M$ in $\msun$, the half-mass radius $\rh$ in pc, the number of black holes contained 
in the cluster $N_{\rm BH}$ and their total mass $M_{\rm BH}$ in $\msun$: the values provided for these quantities in the first part 
of the table refer to the initial properties of the clusters, the ones in the second part to the properties they have at an age of 13 Gyr. 
Moreover, we also provide the total number of black holes contained in the clusters before taking into account the effect of the kick 
velocity, $N_{\rm BH,created}$.}
\label{tab:NBody}
\begin{tabular}{ccccccccccccc}
\hline
$N$-body & \multicolumn{6}{c}{Initial properties}                                                      & & \multicolumn{5}{c}{Final properties} \\
model    & $N$      & $M$                 & $\rh$ & $N_{\rm BH}$ & $M_{\rm BH}$ & $N_{\rm BH,created}$ & & $N$                 & $M$                 & $\rh$  & $N_{\rm BH}$ & $M_{\rm BH}$ \\
\hline
N0       & $10^{5}$ & $5.4 \times 10^{4}$ & $7.6$ & $0$          & $0$          & $176$                & & $8.8 \times 10^{4}$ & $3.2 \times 10^{4}$ & $13.6$  & $0$          & $0$     \\
N0.5     & $10^{5}$ & $6.3 \times 10^{4}$ & $7.6$ & $105$        & $1442$       & $177$                & & $8.5 \times 10^{4}$ & $3.1 \times 10^{4}$ & $14.1$ & $64$         & $486.6$ \\
N1       & $10^{5}$ & $6.3 \times 10^{4}$ & $7.6$ & $176$        & $2024$       & $176$                & & $8.3 \times 10^{4}$ & $3.1 \times 10^{4}$ & $20.0$ &$120$         & $840.3$ \\
\hline
\end{tabular}
\end{table*}

\section{Description of the $N$-body models}
\label{sec:Nbody}

We run three numerical simulations with the $N$-body integrator \textsc{NBODY6} \citep{2003gnbs.book.....A}, in the variant with \textsc{\small GPU} support \citep{2012MNRAS.424..545N}. In the following, we describe the steps we carried out to set up the initial conditions for the simulations and we illustrate their basic properties.

The current properties of NGC 6101 as measured by D15 are presented in the first line of Table~\ref{tab:initialconditions}. We adopted a mass-to-light ratio of $\Upsilon_{\rm V} = 1.9 \, \msun / \lsun$ \citep{2005ApJS..161..304M}. Combined with the $V$-band luminosity of $L_{\rm V} = 5.7 \times 10^{4} \lsun$, we find a present-day mass of $M_{\rm cl} = 1.1 \times 10^{5} \msun$. We then determined the current number of objects in NGC 6101 to be $N = 2.71 \times 10^{5}$ by assuming an average mass of $\bar{m} = 0.4 \msun$. 

\subsection{Estimating the initial conditions of NGC 6101 }
\label{sec:emacss}
To estimate the initial conditions of NGC 6101, we used the fast star cluster evolution code \textsc{emacss} \citep[Evolve Me A Cluster of StarS;][]{2014MNRAS.442.1265A}. By applying H{\'e}non's predictions \citep{1961AnAp...24..369H, 1965AnAp...28...62H} that in a state of balanced evolution, the flow of energy within a cluster is independent of the actual energy source in the core, \textsc{emacss} calculates the evolution of some of its fundamental properties, such as mass, half-mass radius and mean mass. 

We approximate the Milky Way potential by a singular isothermal sphere with $V_{\rm circ} = 220 \, \rm km/s$. We assume a circular orbit at NGC 6101 current Galactocentric radius of $R_{\rm G} = 11.2 \kpc$. For the cluster itself, we assumed a \cite{2001MNRAS.322..231K} initial mass function (IMF) between $0.1$ and $100 \msun$. 

We ran \textsc{emacss} for a grid of different initial values for the total number of objects $N$ and the half-mass radius. Each cluster is evolved to $13 \gyr$ and the half-mass radius and the total mass of the cluster at this age are compared to the current properties of NGC 6101. The initial half-mass radius and total mass of the cluster that provide the best match to the present day properties are chosen as initial conditions of the $N$-body simulations and are given in the second line of Table \ref{tab:initialconditions}. 

We point out that the set of initial properties that we identified in this way correspond to a cluster with an initial half-mass relaxation time of $\trhz = 2.8 \gyr$, implying that NGC 6101 is about $\approx 4.6 \trhz$ old.

\subsection{Model scaling}

Although it is feasible to model NGC 6101 with a direct $N$-body model (see \citealt{2014MNRAS.445.3435H} and \citealt{2016MNRAS.tmp...74W}), we decide to model NGC 6101 with scaled $N$-body models in order to explore several scenarios for the retention of the stellar-mass BHs. We used the approach from \cite{2008MNRAS.389.1858H}, where the scaled model has the same half-mass relaxation time as the real cluster, accounting for the fact that much of the cluster dynamics is dominated by two-body relaxation. If the number of stars in the scaled model is $N^{*}$, then we can find the half-mass radius of the scaled model, $\rh^{*}$, in terms of $N$ and $\rh$ from the expression of $\trh$ (\citealt{1971ApJ...164..399S}, eq. 5):
\begin{equation}
\frac{r_{\rm h}^{*}}{r_{\rm h}}=\left(\frac{N}{N^{*}}\right)^{1/3}\left(\frac{\log\gamma N^{*}}{\log\gamma N}\right)^{2/3}.
\label{eq:scaling_rh}
\end{equation}
Here we set $\gamma = 0.02$ \citep{1996MNRAS.279.1037G}. 

The scaling for the Jacobi radius ($\rj$) is the same as for $\rh$. Because we want to simulate the cluster in the same tidal field strength as for the \textsc{emacss} runs (as mentioned in Section~\ref{sec:emacss}), we need to scale $R_{\rm G}$. The Jacobi radius can be estimated as \citep{1962AJ.....67..471K}:
\begin{equation}
r_{\rm J}=\left(\frac{G M_{\rm cl}}{2\Omega^{2}}\right)^{1/3} \ ,
\label{eq:r_J}
\end{equation}
where $\Omega={V_{\rm circ}}/{R_{\rm G}}$ is the local angular velocity. We can now express the scaling relation for $R_{\rm G}$ as
\begin{equation}
\frac{R_{\rm G}^{*}}{R_{\rm G}}=\left(\frac{N}{N^{*}}\right)^{1/2}\left(\frac{r_{\rm h}^{*}}{r_{\rm h}}\right)^{3/2} \ .
\end{equation}

When making use of scaled $N$-body models, one needs to be aware that processes that have a different $N$-dependence as the relaxation time, are not modelled correctly. We are mostly concerned about the behaviour of the BH population. \cite{2013MNRAS.432.2779B,2013MNRAS.436..584B} showed that the escape rate of BHs in GCs is set by $\trh$ of the cluster as a whole, i.e. not by the time-scale of the BH sub-cluster itself. The fraction of BHs that is retained at an age of 13 Gyr after dynamical evolution should therefore not be affected by the scaling.

For our simulations, we set the number of initial stars to $N^* = 10^5$ and scaled the other properties accordingly (see third line of Table~\ref{tab:initialconditions}).

\subsection{$N$-body simulations}

Because previous works (see discussion in Section \ref{Intro}) have shown that a population of stellar-mass BHs could give rise to a large (observed) core radius, we set up three $N$-body simulations, each of which is characterized by a different fraction of BHs retained with respect to their initial number: in model N1, all the BHs are retained in the cluster, in model N0.5, only 50\% of BHs are retained and in model N0, no BHs are retained.

As initial condition for the three simulations, we consider a set of stars distributed according to the Plummer model \citep{1911MNRAS..71..460P}. We do not include primordial binaries, primarily to speed up the computations \citep[see the discussion in][]{2015MNRAS.450.4070W}. Excluding them may affect the efficiency of BH binary formation and ejection \citep[see][]{2016arXiv160300884C,2016PhRvD..93h4029R}, but because we are mainly interested in studying the difference between clusters with and without BHs, our approach is justified. We adopt the same stellar IMF as for the \textsc{EMACSS} models (Section \ref{sec:emacss}) and we control the removal of the BHs created in the cluster by varying their initial supernova kick velocity. For simulation N1, we set the initial kick velocity to zero, so that all the BHs are kept in the cluster. In the case of simulation N0.5, we want to retain 50\% of the BHs: to do this, for each BH, we draw a random number from a flat distribution in the range $(0,1)$, and we assign a kick velocity greater than the escape velocity to the BH only if the drawn value is above 0.5. This procedure allows us to retain, on average, 50\% of the BHs, without the need of knowing their total number. For simulation N0, we assign a kick velocity greater than the escape velocity to all BHs so that none of them are retained in the cluster. 

The cluster is moving on a circular orbit in the ($x$, $y$)$-$plane with orbital velocity of $V_{\rm circ} = 220 \, \rm km/s$. The stars are evolved with the stellar evolution prescription of \cite{2000MNRAS.315..543H} for a metallicity of $\feh = -1.98$. We summarize the initial and final (after 13 Gyr) properties of the simulations in Table~\ref{tab:NBody}. 

\begin{figure*}
\includegraphics[width=1\textwidth]{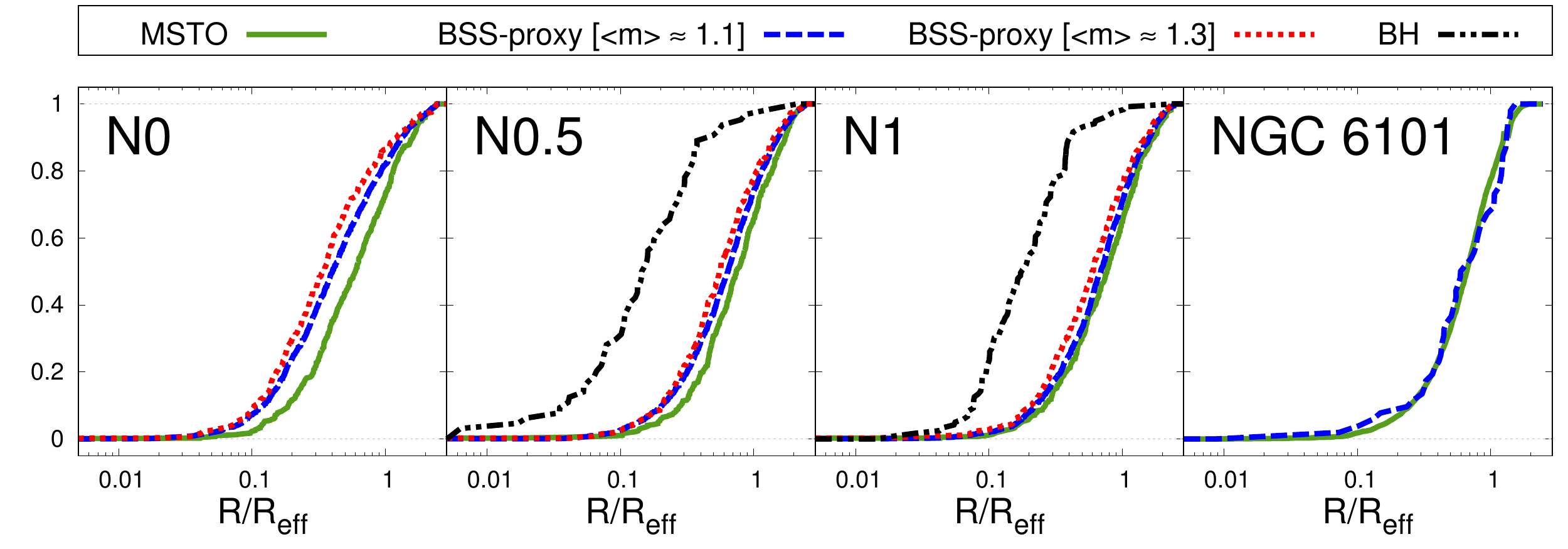}
\caption{Cumulative radial distribution of different groups of stars. In the first three panels, we show the distributions of the MSTO stars (solid green lines), of the two samples of BSS-proxy stars (the one with average mass of $\langle m \rangle = 1.1\msun$ is represented with dashed blue lines, the one with $\langle m \rangle = 1.3\msun$ with dotted red lines), and the black holes (double dot$-$dashed black lines) as a function of the projected distance from the centre, in units of the projected effective radius $R_{\rm eff}$. Each panel corresponds to the $13 \gyr$ snapshot of a different simulation, as indicated by the labels. For comparison, in the last panel on the right, we provide a copy of fig. 7 of D15 with the measured cumulative radial distributions of BSSs and MSTO stars in NGC 6101. \label{fig:CPlot-All-1} }
\end{figure*}

\section{Results}
\label{sec:Results}

With the $N$-body models of NGC 6101 in place, we perform the same analysis as carried out by D15. We compare the cumulative radial distribution for the different star types in our numerical simulations to the one presented in D15. Next, we analyse the MF slope of our $N$-body models and again compare them to results presented by D15 for NGC 6101. We could, in principle, also analyse the radial distribution of the binary stars, but the models were created without primordial binaries, and the binaries that formed in the course of the simulation are too few to give a meaningful result.

\subsection{Cumulative radial distribution}

In their fig. 7, D15 show the cumulative radial distribution of four different groups of stars observed in NGC 6101: BSSs, horizontal branch stars, red giant branch stars (RGBs) and MSTO stars. They point out that the four groups have the same distribution in the cluster. Here we consider the same quantities in our simulations, and compare them with their finding.

We focus our analysis on the distribution of BSS and MSTO stars, because they have the largest mass difference among the star types analysed by D15 and, if the cluster is mass-segregated, should therefore have the largest difference in spatial distribution. We label stars in the mass range $0.79-0.81 \msun$ as MSTO stars. 

No BSSs were created in our simulations, because we did not include primordial binaries in our simulations: it has been shown in observational studies \citep{2008A&A...481..701S,2009Natur.457..288K} as well as in simulations \citep{2013ApJ...777..106C,2013ApJ...777..105S} that the efficiency of BSS formation is positively correlated with the binary fraction. The number of binaries has not only an influence on the BSSs created by mass transfer in a binary but also an important influence on the BSSs created by collisions, as the majority of these collisions are binary-mediated. So the number of primordial binaries directly affects the creation of BSSs. 

We therefore need a proxy for the BSSs and since the only relevant property for this analysis is the mass of the stars and not their type, we use white dwarfs (WDs) as a proxy for BSSs. WDs are the only abundant objects for which the mass range reaches values high enough to be comparable to BSSs. Red giants, for example, which are the evolved stars\footnote{In this work, every post-main-sequence (MS) star which is not a remnant is regarded as an evolved star.} with the highest mass in our simulations only reach slightly above the MSTO mass after $13 \gyr$ of evolution and are therefore not a good BSS proxy candidate. 

The estimated mass range for BSSs in GCs ranges from $0.6$ to $3.74 \msun$ \citep{1997ApJ...489L..59S,1998ApJ...507..818G,2005ApJ...632..894D,2012ApJ...754...91L,2014ApJ...783...34F} with average mass in the range between $1.0$ and $1.3 \msun$ \citep{2005ApJ...632..894D,2012ApJ...754...91L}. With particular reference to the GC NGC 6101, the only available information on the mass of its BSSs is that for all of them, it holds that $M_{\rm BSS} \leq 2 \, M_{\rm MSTO} \approx 1.6 \msun$ \citep{2001A&A...380..478M}. Because the determination of masses of BSSs is very uncertain, we decide to consider two different mass ranges for the BSS proxies in our simulations: the first sample is formed by WDs with masses in the range of $1.0-1.5 \msun$ and an average mass of $1.1 \msun$, the second sample by WDs with masses in the range of $1.187-1.5 \msun$ and an average mass of $1.3 \msun$. We note here that the largest mass for a WD in our simulations is $1.5 \msun$.

\begin{figure}
\includegraphics[width=0.95\columnwidth]{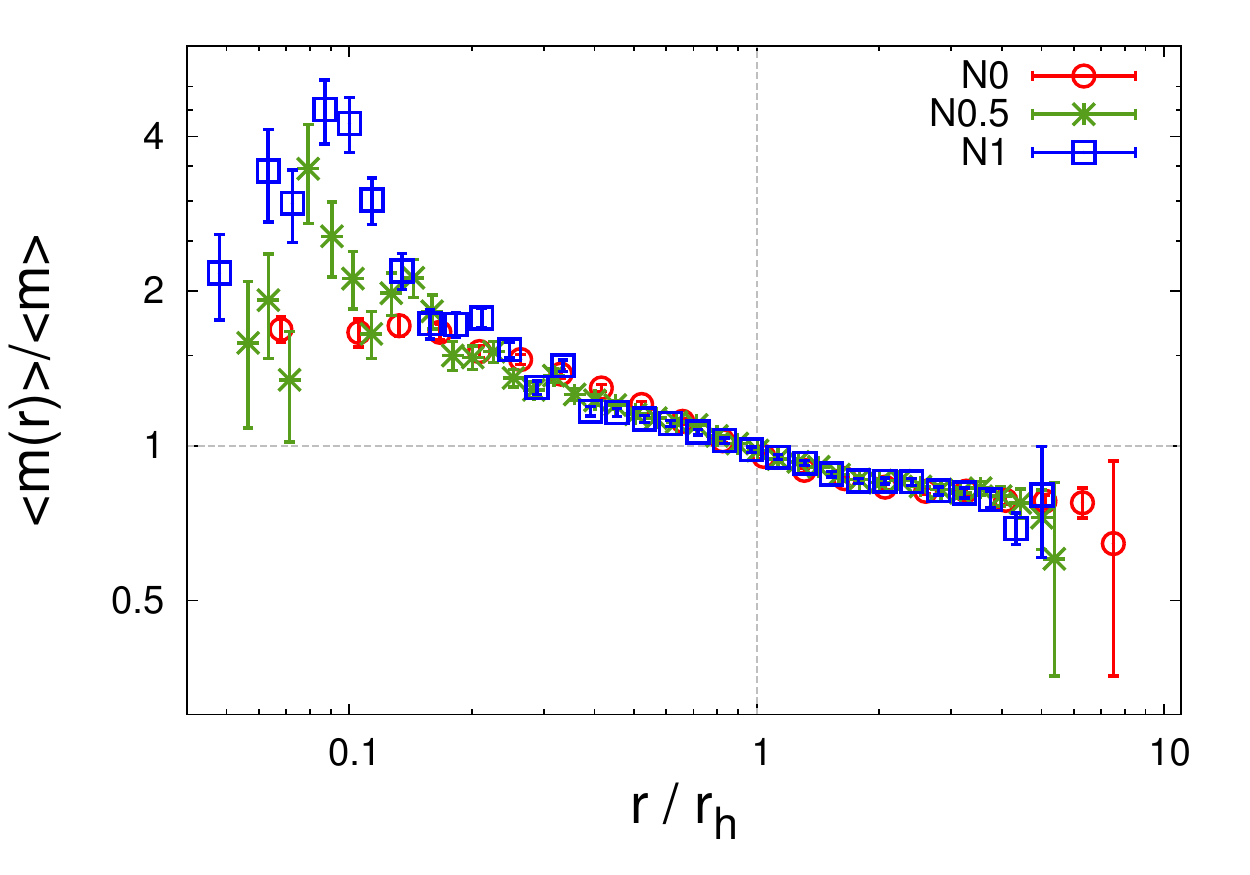} 
\caption{Relative mean mass as a function of the distance from the cluster centre in units of $\rh$. The relative mean mass corresponds to the ratio of the mean mass of stars in radial bins divided by the total mean mass. Circles (red), stars (green) and boxes (blue) refer to simulation N0, N0.5 and N1, respectively. Error bars denote $1\sigma$ uncertainties.
\label{fig:support}}
\end{figure}

Using WDs as a proxy for BSSs may overestimate the degree of central concentration. This is because WDs were more massive in the recent past and therefore could still be migrating outwards towards their relaxed position \citep[for an observational study of this process, see][]{2013ApJ...778..104R}. Some of the BSSs (in particular, the ones that are formed from the merger of stars induced by collisions) on the other hand could still be migrating inwards because they were less massive in the recent past. The amount of mass segregation that is inferred from the comparison of the distribution of WD and MSTO stars in our simulations is therefore overestimated with respect to the one that could be obtained when considering BSSs.

In Fig.~\ref{fig:CPlot-All-1}, we show the cumulative distribution of projected distances for the MSTO stars, the BHs and the two BSS-proxies samples for the three simulations. In all cases, the cluster is projected along the $z$-axis (we carefully checked that the results do not depend on the choice of the selected projection axis). The result of D15 for NGC 6101 is presented in the right-hand panel of Fig.~\ref{fig:CPlot-All-1}. With an increasing number of BHs retained in the cluster, the differences in radial distribution of MSTO stars and BSS-proxies diminish. The different amount of mass segregation observed in this way is surprising at first sight because all three models have evolved for the same amount of dynamical time. In the next sections, to further understand this issue, we proceed by analysing the mass distribution of all the objects, including the remnants [WDs, neutron stars (NSs) and BHs].

\subsection{Mean mass at different radii}
\label{sub:meanmass}

In Fig.~\ref{fig:support}, we show the relative mean mass, i.e. the mean mass of all objects in radial bins divided by the global mean mass, as a function of the distance from the cluster centre in units of the half-mass radius, for our three simulations at $13 \gyr$. By comparing the relative mean mass of the three clusters, we see that they show the same behaviour. The main difference is found in the innermost region ($r/r_{\rm h} \lesssim 0.1 $) where, in the snapshots with BHs, the scatter around the common mean value is greater than in the snapshot without BHs.

\begin{figure}
\includegraphics[width=0.95\columnwidth]{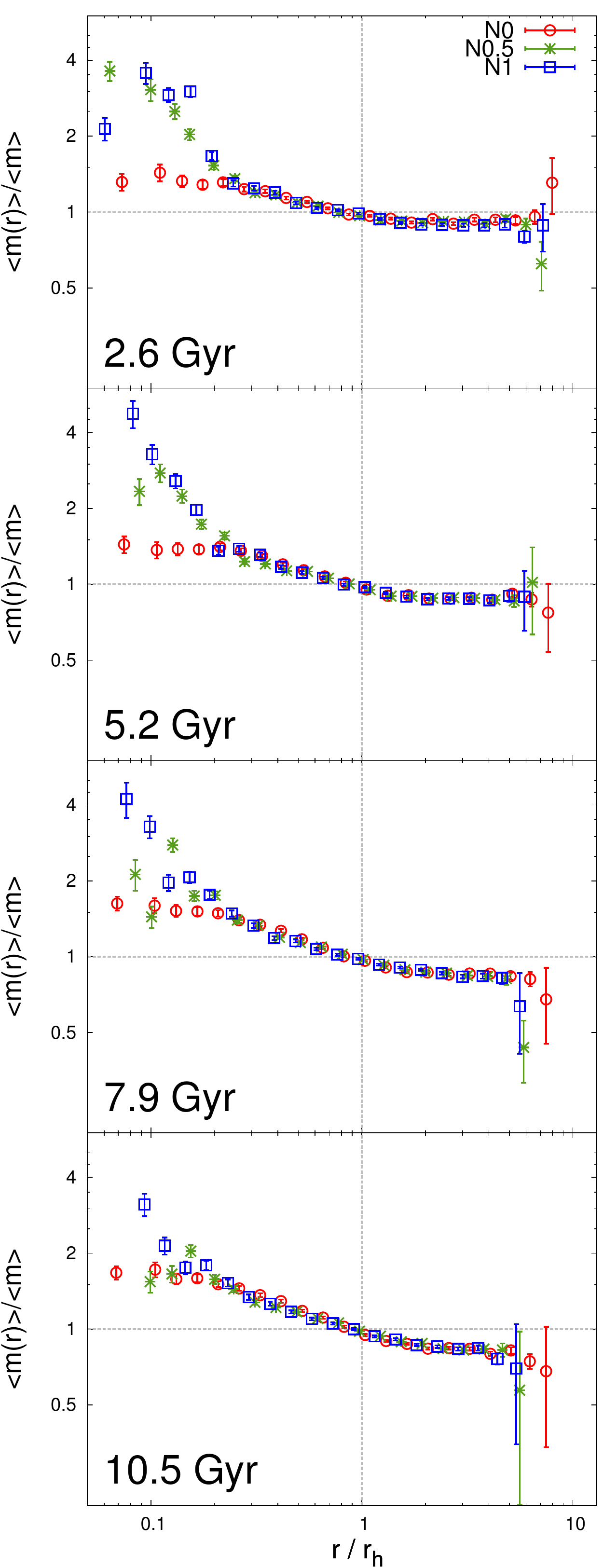}
\caption{Relative mean mass (i.e. mean mass of stars in radial bins divided by the total mean mass) as a function of the distance from the cluster centre in units of $\rh$, for all simulations at four different times: $2.6$, $5.3$, $7.9$ and $10.5 \gyr$. Circles (red), stars (green) and boxes (blue) refer to simulation N0, N0.5 and N1, respectively. Errorbars denote $1\sigma$ uncertainties.
\label{fig:AVG-Plot-Selection}}
\end{figure}

We analyse the radial dependence of the relative mean mass at different times, and we find that it has the same overall behaviour independently of the presence of BHs. Only in the inner regions ($r/r_{\rm h} \lesssim 0.1 $), there is a difference between the profiles of different simulations at different times, with the mean mass in the centre increasing faster in the models with BHs. In Fig.~\ref{fig:AVG-Plot-Selection}, we show the relative mean mass for the three simulations at four different times in their evolution ($2.6$, $5.3$, $7.9$ and $10.5 \gyr$). A change in the slope of this quantity is observed for all the simulations: outside the half-mass radius, it becomes steeper in time. For single-mass systems, it is well understood that after several relaxation times the evolution is self-similar \citep{1961AnAp...24..369H,1965AnAp...28...62H}. No studies regarding the evolution of the structure of multimass systems exist yet, but it has been shown that the evolution of mass and radii of multimass models is comparable to the single-mass case \citep{1995ApJ...443..109L,2010MNRAS.408L..16G}, but faster in time. These $N$-body results suggest that there also exists self-similarity in terms of the mean mass profile.

We find that the regions where the BSS-proxies are located are significantly smaller in the cases with BHs, and are less central than in the case without BHs, as show in Fig.~\ref{fig:CPlot-All-1}. When BHs are present in the system, we find that the rest of the stars are pushed outwards, and their distributions are more similar to one another. Multimass collisional systems will try to reach equipartition, but the lowest mass stars never reach equipartition because they are in the regime near the truncation energy \citep{1981AJ.....86..318M,2006MNRAS.366..227M,2015arXiv150802120G,2016arXiv160300878B}. \cite{2016arXiv160300878B} use Monte Carlo simulations to show that clusters only achieve partial equipartition, and that stars with masses below $m_{\rm eq}$ have similar velocity dispersions, independent of mass, while stars with $m >> m_{\rm eq}$ achieve equipartition ($\sigma \sim m^{-1/2}$). These authors show that $m_{\rm eq}$ is larger for low-concentration models. For models with a wide mass spectrum, $m_{\rm eq}$ is also higher (comparable to the density-weighted mean mass in the core), and for clusters containing BHs, this could be above the turn-off mass. This implies that the absence of mass segregation among visible stars can be a signal that $m_{\rm eq}$ is much higher than the turn-off mass: at the age of GCs, this can only be due to BHs.

\begin{figure}
\includegraphics[width=0.95\columnwidth]{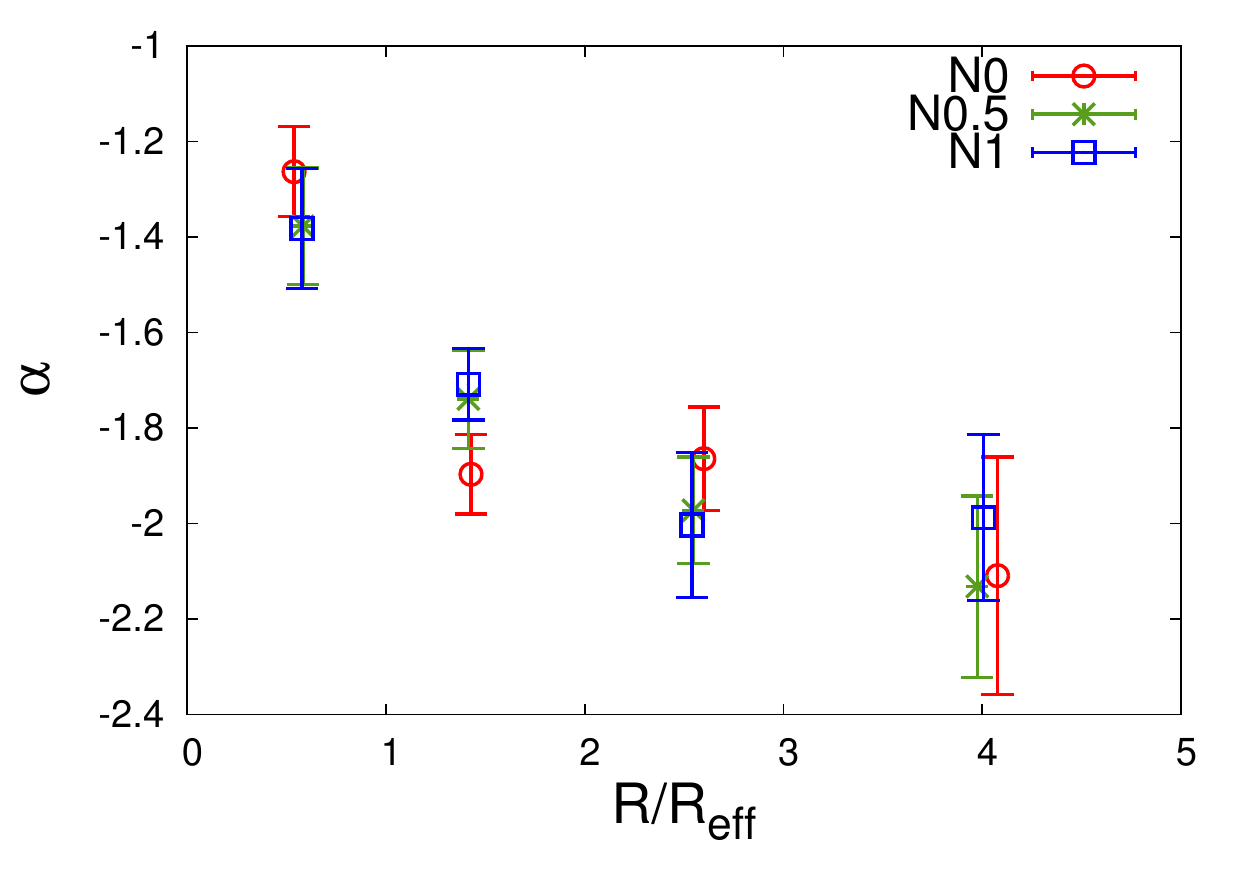}
\caption{Mass function slope ($\alpha$) as a function of the projected distance from the centre in units of $R_{\rm eff}$ for all three simulations at $13 \gyr$. Circles (red), stars (green) and boxes (blue) refer to simulation N0, N0.5 and N1, respectively. Error bars denote $1\sigma$ uncertainties.
\label{fig:MF_plot_1}}
\end{figure}

\begin{table*}
\caption{Main properties of the single-mass model SM, and of the multimass models M0, M0.5 and M1, as listed in the first column. We provide the total mass of the system $M$, the total mass of black holes $M_{\rm BH}$ and the mean mass of black holes $\langle m_{\rm BH} \rangle$, all expressed in $\msun$, the half-mass radius $\rh$, the projected effective radius $R_{\rm eff}$ and the assumed Jacobi radius $r_{\rm J}$ in pc and the central line-of-sight velocity dispersion $\sigma_0$ in km/s. The last two columns of the table refer to the structural parameters of the models that were determined through the fitting procedure (errors are also listed), namely the concentration parameter $W_0$ and the  truncation parameter $g$.}
\label{tab:Models}
\begin{tabular}{cccccccccc}
\hline
Model & $M$                & $M_{\rm BH}$ & $\langle m_{\rm BH} \rangle$ & $r_{\rm h}$ & $R_{\rm eff}$ & $r_{\rm J}$ & $\sigma_0$ & $W_0$               & $g$                \\
\hline
SM    & $1.08\times10^{5}$ & $0$          & $0$                          & $12.4$      &  $9.3$        & $84.6$      & $2.7$      & $6.0_{-0.9}^{+0.8}$ & $1.0_{-0.4}^{+0.5}$   \\ 
M0    & $1.08\times10^{5}$ & $0$          & $0$                          & $16.8$      &  $12.6$       &  $84.6$     & $2.2$      & $7.5_{-1.4}^{+1.1}$ & $0.8_{-0.39}^{+0.41}$          \\
M0.5  & $1.14\times10^{5}$ & $1770$       & $7.6$                        & $11.6$      &  $8.7$        & $86.0$      & $3.0$      & $28_{-12}^{+8.4}$   & $2.08_{-0.07}^{+0.09}$ \\
M1    & $1.20\times10^{5}$ & $3225$       & $7.0$                        & $11.4$      &  $8.6$        & $87.4$      & $3.2$      & $22_{-9.1}^{+9.2}$  & $2.13_{-0.06}^{+0.07}$ \\
\hline
\end{tabular}
\end{table*}

\subsection{MF slope}

Another indication that NGC 6101 lacks mass segregation is the observation that the cluster MF slope is independent of the distance to the cluster centre (D15). In a mass-segregated cluster, one expects a radius-dependent MF slope \citep[e.g.][]{2014MNRAS.442.1569W}; we therefore need to also study whether a stellar-mass BH population can reproduce a radius-independent constant MF slope as seen in NGC 6101 to further confirm our theory. 

To study the slope of the MF, we used a procedure similar to the one used by D15. First, we project the $N$-body data from each model at $13 \gyr$ along the $z$-axis and then we select the MS and the evolved stars in the mass range $0.35-0.7 \msun$, which corresponds to the mass range of stars in the FORS2 data set used by D15. Next, we divide the stars in four concentric annuli ($0.0-1.0 \, R_{\rm eff}$, $1.0-2.0 \, R_{\rm eff}$, $2.0-3.5 \, R_{\rm eff}$ and $3.5-5.0 \, R_{\rm eff}$) and further separate them in 10 mass bins. Finally, for each annulus, we fit a power law of the form ${\rm d}N/{\rm d}m \sim m^{\alpha}$, to the mass bins and determine the MF slope.

In Fig.~\ref{fig:MF_plot_1}, we show the MF slope of our simulations as a function of projected radius. For the simulation without BHs, we see a decrease of the MF slope with increasing distance from the cluster centre. With increasing amount of initially retained BHs, the MF slope becomes flatter. The MF of the N1 simulation varies negligibly with radius and therefore we can reproduce the radius-independent MF slope as found in NGC 6101 with our $N$-body model which initially retained all BHs.

\begin{figure}
\includegraphics[width=1\columnwidth]{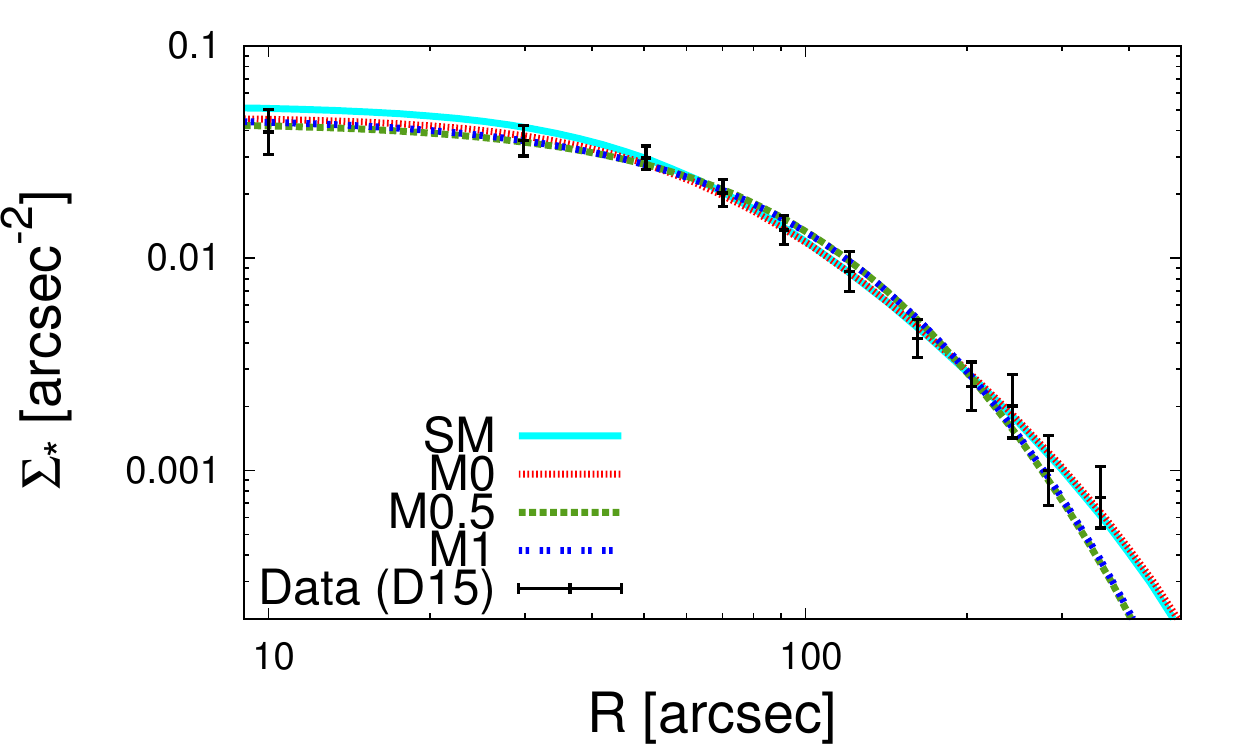} 
\caption{Number density profile of the globular cluster NGC 6101. Black points with error bars indicate the measurements from D15. Solid (cyan), fine dotted (red), thick dotted (green) and double dotted (blue) lines represent the profiles of the best-fitting models SM, M0, M0.5 and M1, respectively. Error bars denote $1\sigma$ uncertainties.
\label{fig:nd_profile}}
\end{figure}

\begin{figure*}
\includegraphics[width=0.75\textwidth]{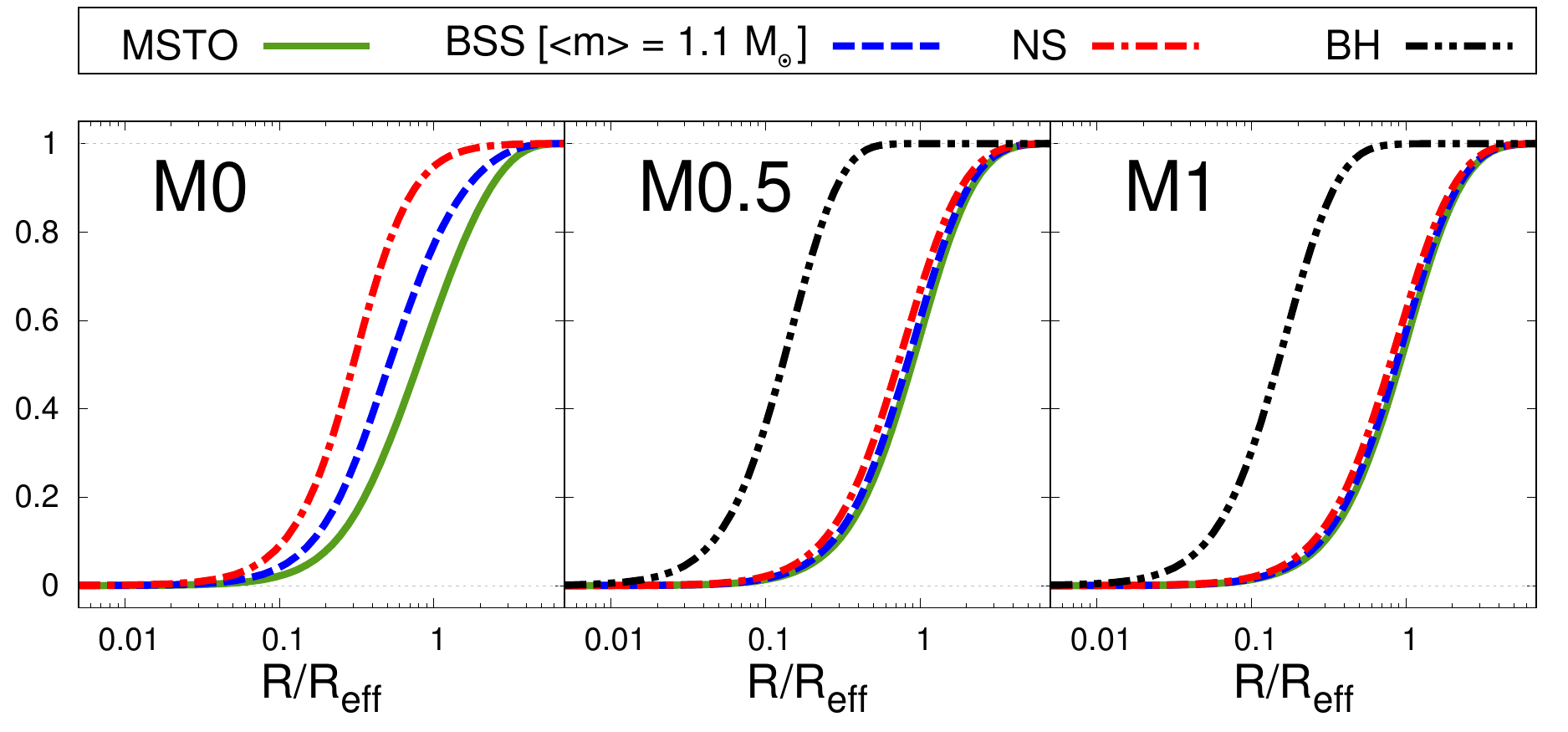}
\caption{Cumulative radial distribution of different groups of stars. We show the distributions of the MSTO stars (solid green lines), of the BSS-proxy stars (dashed blue lines), of the NSs (double dash$-$dotted red lines) and of black holes (double dot$-$dashed black lines) as a function of the projected distance from the centre, in units of the projected effective radius $R_{\rm eff}$. Each panel corresponds to one of the multimass models we considered, as indicated by the labels. The average mass for the MSTO stars is $0.8 \msun$, for the NSs $1.4 \msun$ and for the BSS-proxy stars $1.1 \msun$. For the BHs, the average mass is $7.6 \msun$ for M0.5 and $7.0 \msun$ for M1.
\label{fig:model_cum_profile} }
\end{figure*}

\section{Prediction for the expected kinematics}
\label{sec:discussion}

The observable properties that are often used to quantify the amount of mass segregation in GCs are the cumulative radial distribution of their observable stars, or the MF slope at different radii. In Section \ref{sec:Results}, we showed that, by using these observations, it is impossible to distinguish between a cluster that is not mass-segregated, and a mass-segregated cluster containing a population of stellar-mass BHs. Moreover, a cluster containing an IMBH would also show no sign of mass segregation as the IMBH halts the mass segregation process \citep{2008ApJ...686..303G}. It is therefore necessary to identify an additional observational property to discriminate between these three options. 

In addition, we note that the two $N$-body models with BHs can reproduce the large core and the missing observable mass segregation, while the one without BHs is not able to reproduce either. From these models, it is therefore not possible to conclude whether the absence of observable mass segregation is due to the BHs, or due to the large core. \cite{2016arXiv160300878B} found that $m_{\rm eq}$ (i.e. the mass below which stars have similar distributions, see discussion in Section \ref{sub:meanmass}) depends on the concentration of the cluster. To make sure that the absence of mass segregation is not due to the large core, which could be the result of other physics that was not included in our $N$-body models (i.e. a high primordial binary fraction; \citealt{1994ApJ...431..231V}, \citealt{2011MNRAS.410.2698G},  or an even larger core at formation), we consider equilibrium models that are able to include different mass components (so-called multimass models) and the effect of mass segregation. In these models, we can vary the stellar MF, and adjust the central concentration to match the observed number density profile of NGC 6101, to take advantage of their predictive power. 

We use the models which are provided by the software package {\sc limepy}\footnote{{\sc limepy} can be found at: \url{https://github.com/mgieles/limepy}} \citep[Lowered Isothermal Model Explorer in Python;][]{2015arXiv150802120G}. This package allows the user to compute models including multiple mass components, and a variable amount of radial anisotropy. These models are a solution to the collisionless Boltzmann equation assuming a Maxwellian velocity distribution that is `lowered' to mimic the effect of an escape energy due to the Galactic tides. {\sc limepy} models include the well-known single-mass King model \citep{1966AJ.....71...64K} as well as its multimass extension by \cite{1976ApJ...206..128D}. The models include the truncation prescription by \cite{2014JSMTE..04..006G} with a continuous truncation parameter allowing us to model clusters in between the three classically known \cite{1954MNRAS.114..191W}, \cite{1966AJ.....71...64K} and \cite{1975AJ.....80..175W} models. The {\sc limepy} models accurately describe the phase-space density of $N$-body models of single-mass systems \citep{2016arXiv160502032Z} and multimass systems (Peuten et al., in preparation).

We consider four different dynamical models, to take into account the different scenarios introduced above: 
\begin{itemize}
\item a single-mass model, representing a cluster with no mass-segregation; in the following, we refer to this as model SM. 
\item a multimass model representing a mass segregated cluster containing no BHs. Since it is comparable to $N$-body simulation N0, we name it M0.
\item a multimass model, identified as model M0.5, representing a mass-segregated cluster containing 36\% of its initial population of BHs, corresponding to the number of BHs retained in model N0.5 after $13 \gyr$.
\item a multimass model, called M1, representing a mass-segregated cluster containing 68\% of its initial population of BHs, corresponding to the number of BHs retained in model N1 after $13 \gyr$.
\end{itemize}

To calculate the multimass models M0, M0.5 and M1, it is necessary to provide an MF. Here we use the MF determined from the $N$-body models, by considering five mass bins for the MS stars, three mass bins for the WDs and one bin each for the evolved stars, NSs and BHs (for a discussion about the selection of mass bins, we refer the reader to Peuten et al., in preparation). 

We summarize the properties of all models in Table~\ref{tab:Models}. For all models, the luminosity of the cluster is set to the value given in the first line of Table~\ref{tab:initialconditions}. 
The mass is determined by assuming the same mass-to-light ratio as in Section~\ref{sec:Nbody} with value of $\Upsilon_{\rm V} = 1.9 \, \msun / \lsun$ for models SM and M0. To account for the mass of the BHs that do not contribute to the luminosity of the cluster, we changed the mass-to-light ratio to $\Upsilon_{\rm V} = 2.0 \, \msun / \lsun$ for M0.5 and $\Upsilon_{\rm V} = 2.1 \, \msun / \lsun$ for M1 to calculate their respective mass. For each model, we calculated the expected Jacobi radius using equation (\ref{eq:r_J}) with the same assumptions as in Section~\ref{sec:Nbody} and the above estimated mass. With these choices, we set the scales of the models, and we are only left with two structural parameters and a scale to fit on: $W_0$, which determines the concentration of the models, $g$, the truncation parameter and an additional normalization parameter which accounts for the unknown number of total stars used in the number density profile by D15 and which has no physical meaning to the results.

We carry out the fits by means of the \textsc{emcee} \citep{2013PASP..125..306F} software, which is a pure-\textsc{python} implementation of the Goodman \& Weare's Affine Invariant Markov chain Monte Carlo Ensemble sampler \citep{10.2140/camcos.2010.5.65}. For the case of the multimass models, we fit on the number density of the evolved stars as these are the stars for which it is possible to obtain measurements. The best-fitting parameters obtained with this fitting procedure are provided in Table~\ref{tab:Models}.

For the analysis of the case with an IMBH, we considered the family of dynamical models\footnote{Several pre-tabulated models from this family are available for download at: \url{http://www.cosmic-lab.eu/bhking/}} presented by \cite{2007MNRAS.381..103M}, which describe a single-mass cluster with an IMBH in the centre. Within this family, we selected the model characterized by $W_0 = 7.75$ and $M_{\rm IMBH} / M = 0.01$. We chose this model because it resembles the number density profile of NGC 6101, and its IMBH mass is comparable to the total mass of the BH population in model M0.5. To scale this model, we use the same scales we assumed for model M0.5.

In Fig.~\ref{fig:nd_profile}, we show the number density profiles of the four best-fitting dynamical models together with the observed values from D15. All the models appear to describe the observed data  well. A slight disagreement is only seen at large radii, where the two dynamical models including BHs slightly underestimate the outermost point. For the three multimass models, we also calculated the cumulative radial distributions of MSTO stars, BSSs, NSs and BHs. As proxy for the BSSs, we chose the WD mass bin with an average mass of $1.1 \msun$ and as MSTO stars, we chose the mass bin with an average mass of $0.8 \msun$. Fig.~\ref{fig:model_cum_profile} shows the same behaviour already found for the numerical simulations, and shown in Fig.~\ref{fig:CPlot-All-1}: when considering a larger number of BHs in the cluster, the distributions of the other types of stars become more similar.

\begin{figure}
\includegraphics[width=1\columnwidth]{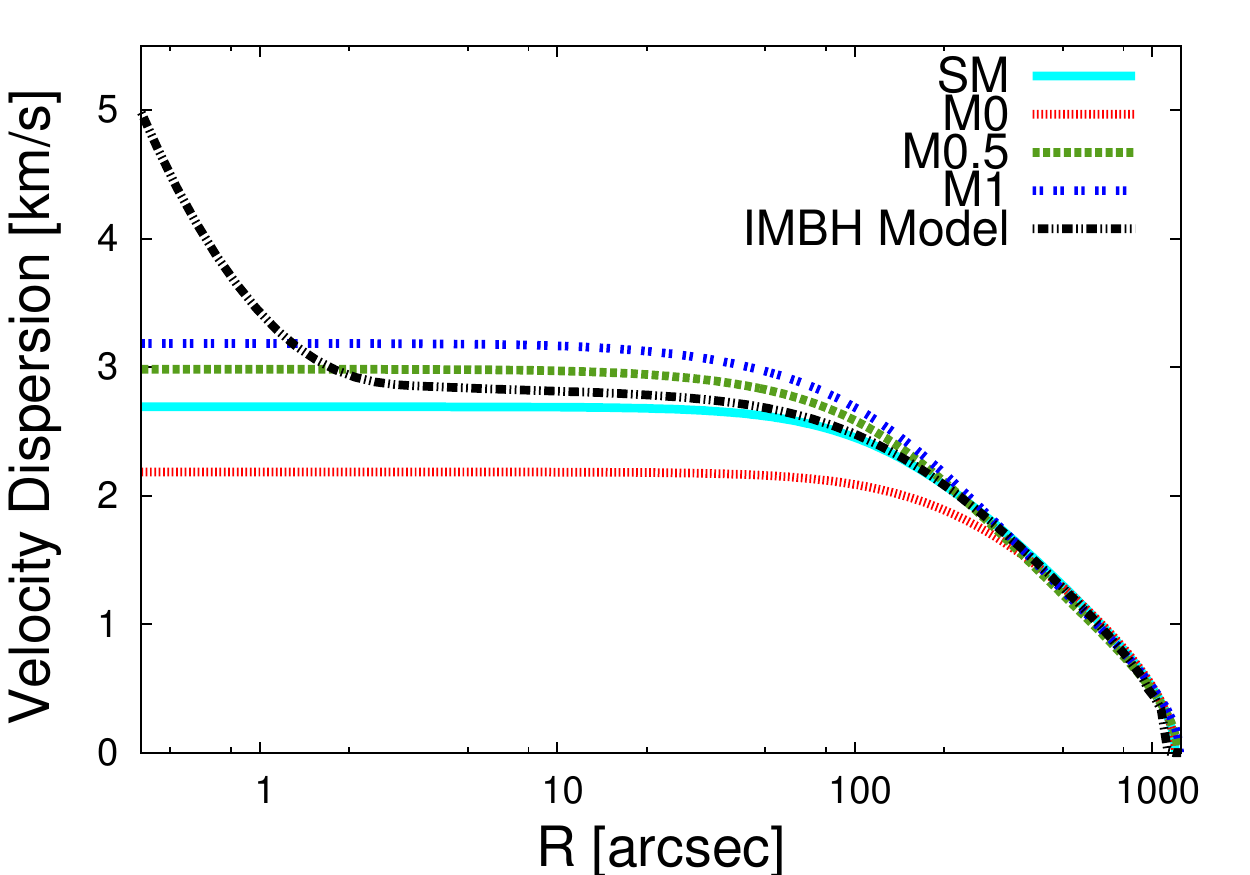} 
\caption{Line-of-sight velocity dispersion profiles predicted by the models. Solid (cyan), fine dotted (red), thick dotted (green), double dotted (blue) and dot$-$dashed (black) lines show the profiles predicted from the SM, M0, M0.5, M1 and IMBH model, respectively. In the case of multimass models, we show here the profiles relative to the mass bin representing evolved stars, which are the ones for which is possible to obtain measurements.
\label{fig:vd_profile}}
\end{figure}

As a way to distinguish between the different proposed explanations for the missing signatures of mass segregation, we consider the line-of-sight velocity dispersion profiles predicted by the models and shown in Fig.~\ref{fig:vd_profile}. For the models with mass segregation, we used again only the evolved stars for the calculation of the line-of-sight velocity dispersion, as these are the stars for which it is possible to obtain measurements. In the $N$-body models, these stars have luminosities in the range of $\log\left({\rm L/{\rm L_{\sun}}}\right)=0.7-3.3$. For the models without mass segregation, a selection of stars is irrelevant as the line-of-sight velocity dispersion is the same for all stars. The values of the central line-of-sight velocity dispersion obtained for models M0.5 and M1, representing mass-segregated clusters with BHs, are respectively $0.3$ and $0.5 \, \rm {km \, s^{-1}}$ larger than the one obtained for model SM (no mass segregation, no BHs), which is, in turn, $0.5 \, \rm{km \, s^{-1}}$ larger than that predicted for model M0 (mass-segregated cluster without BHs). The line-of-sight velocity dispersion profile for the IMBH model follows the SM profile. It only differs in the central $10 \, \rm arcsec$, where a cusp in the velocity profile is found. We note that the number density profile does not resolve this area, and therefore the central rise expected when an IMBH is present cannot be detected. The central line-of-sight velocity dispersion for the IMBH model is $2.4 \, \rm {km \, s^{-1}}$ larger than the one of model SM: this means that a measurement of the line-of-sight velocity dispersion within the inner $10 \, \rm arcsec$ could distinguish between the scenarios.

\section{Discussion and conclusion}
\label{sec:Conclusion}

Recently, D15 observed that BSS and MSTO stars have the same radial distribution in the GC NGC 6101, and they argue that the cluster is not mass-segregated and not dynamically evolved. \cite{1991AJ....102..628S} and \cite{2001A&A...380..478M}, who also studied the radial distribution of the BSSs in this cluster, found indications for mass segregation. The reason for this discrepancy is that each of these papers analyse a different sample of BSS stars. \cite{1991AJ....102..628S} were the first ones to study BSSs in this cluster and they found 28 BSSs. \cite{2001A&A...380..478M} found and studied 73 BSSs in NGC 6101. D15, however, reduced the sample of BSSs in NGC 6101 to 52 objects, after identifying and removing sources which are contaminated and/or blended by other MS stars or evolved stars. Given these and other improvements by D15, we adopt their interpretation that NGC 6101 does not show any observable signs of mass segregation. 

By carrying out three numerical $N$-body simulations containing a different amount of BHs, we showed that the same behaviour is found in a mass-segregated cluster containing a population of stellar-mass BHs. Indeed, even if they are not directly observable, BHs have an effect on the overall distribution of stars in the mass range available for observations ($0.7-1.6 \msun$) that appear to have the same distribution, as shown in Fig.~\ref{fig:CPlot-All-1}.

We also see from our simulation without BHs (N0) that the age and present-day mass and half-mass radius suggest that NGC 6101 is dynamically evolved, and is expected to be mass-segregated. Model N0 shows clear evidence for observable mass segregation. The scenario of a non-mass-segregated cluster could then only be explained if some of our assumptions, such as the age of the cluster, its stellar evolution or its IMF, were significantly different from what we assumed here, which we consider unlikely. We therefore favour the explanation that NGC 6101 contains a stellar-mass BH population.

Stellar-mass BH candidates were recently found in M22 and M63 by \cite{2012Natur.490...71S} and \cite{2013ApJ...777...69C}, respectively. Several studies have shown that, if the initial supernova kicks are not large enough to eject the BHs from the cluster at creation, then a significant fraction of BHs can be retained for more than $12 \gyr$  \citep{2013MNRAS.432.2779B,2013MNRAS.436..584B,2013MNRAS.430L..30S,2015ApJ...800....9M}; in particular, this happens when clusters have large initial radii \citep{2015ApJ...800....9M, 2016arXiv160104227R}. Moreover, \cite{2008MNRAS.386...65M} showed that the large cores of GCs in the Magellanic Clouds can be explained by the presence of a population of stellar-mass BHs in the systems. The fact that NGC 6101 is on a retrograde orbit is seen as an indication for an extragalactic origin \citep{1995AJ....109..605G}. More recently, it has been suggested that NGC 6101 was accreted into the Milky Way \citep{2004MNRAS.355..504M} and could originally come from the Canis Major dwarf galaxy \citep{2004MNRAS.348...12M}. One of the arguments used by \cite{2004MNRAS.355..504M} to support the claim that NGC 6101 is accreted is the observation that the large core radius of the cluster is more comparable to the core radii of GCs in dwarf galaxies than to those of clusters in the Milky Way. This raises the question why GCs that form in dwarf galaxies contain more BHs than GCs that form \textit{in situ}. There are no reasons to expect that the initial stellar MF is significantly different in dwarf galaxies (although, see \citealt{2013ApJ...771...29G}), nor that the supernova kicks are different in dwarf galaxies. One idea is that all GCs retain a large fraction of their BHs after supernova kicks (i.e. BH kicks are low), and that  GCs in dwarf galaxies form with lower densities (e.g. \citealt{2008ApJ...672.1006E}). A low-density implies a long $\trh$ (for a given mass), such that fewer BHs are dynamically ejected.

An alternative explanation for the observed properties of NGC 6101 could be the presence of an IMBH. This central object would cause, in many respects, effects similar to those of a population of stellar-mass BHs, such as the formation of a large core and a large ratio of core radius to half-mass radius \citep{2007MNRAS.374..344T,2013A&A...558A.117L}. In addition, an IMBH can quench mass segregation among the visible stars: \cite{2008ApJ...686..303G} show that this effect is due to close encounters between stars and the IMBH, resulting in slingshot ejections to large distances, thereby reversing mass segregation. Moreover, the IMBH is likely to acquire a companion, either a star or a remnant, which makes stellar ejections particularly common. \cite{2008ApJ...686..303G} also measured mass segregation by looking at the variation with radius of the average mean mass of MS stars with mass in the range $0.2-0.8 \msun$. They showed that an IMBH with mass equal to $1\%$ of the cluster mass generates a small variation of this quantity between the centre and the half-mass radius, and they conclude that if such variation is smaller than $\sim 0.07 \msun$, the cluster is likely to be hosting an IMBH. As a comparison, in both our $N$-body simulations containing a population of BHs, the variation of the average mass between the centre and the half-mass radius is also smaller than $0.07 \msun$ (for N0.5, we find a variation of $0.03 \msun$, for N1 of $0.04 \msun$), when considering MS stars. \cite{2008ApJ...686..303G} also discuss the possibility that a BH population could create the same observable effect, but they assume that BHs will leave the GC rather quickly and their impact on the observed mass segregation should therefore be rather small.

Another possible alternative explanation could be the presence of binaries alone: it is known that binaries inflate the core \citep{1994ApJ...431..231V,2011MNRAS.410.2698G} and could therefore also explain the large core of NGC 6101. With our current results, we cannot quantify the degree of expected mass segregation due to binaries. A qualitative result can be drawn if one assumes that the binaries have a mass distribution comparable to the one of the NSs: the first panel of Fig.~\ref{fig:model_cum_profile}, relative to model M1, shows that the NSs alone have a negligible effect on the apparent observational mass segregation and therefore the expected effect due to binaries alone should also be rather low.

Due to recent confirmation of the existence of gravitational waves by a binary BH merger \citep{2016PhRvL.116f1102A}, it is worth mentioning that GCs with a sizeable BH population, such as NGC 6101, could be a cradle of gravitational wave sources \citep{2000ApJ...528L..17P,2012MNRAS.422..841A}: not only do recent studies show that a significant fraction of BHs can be retained for more than $12 \gyr$ but they also predict a high binary fraction among the BHs in the core \citep{2015ApJ...800....9M}. 

Finally, we propose an observational test to distinguish the various possible scenarios for the cluster. From a comparison of distribution function-based models to the number density profile of NGC 6101, we show that a mass-segregated cluster with stellar-mass BHs is expected to have a central line-of-sight velocity dispersion $\sim 0.5 \, \rm {km \, s^{-1}}$ larger than a non-segregated cluster without BHs. When considering the presence of an IMBH in the centre of the cluster, the predicted central line-of-sight velocity dispersion should be even larger, assuming a value up to $\sim 5.1 \, \rm {km \, s^{-1}}$. Looking at the star counts by D15 for NGC 6101, one can see that approximately 100 RGB stars (or $20\%$ of D15 sample), with a $V$-band magnitude between $13.5$ and $18.7$, are located within the core radius, with around $7$ of them located within the inner $10$ arcsec. By obtaining an accurate measure of the velocity dispersion of NGC 6101 within the core radius, it should be possible to discriminate between the proposed scenarios, and to determine the dynamical state of this cluster.

\section*{Acknowledgements}
We thank the anonymous referee for constructive suggestions. We also thank the organisers of the Third Gaia Challenge Workshop (Barcelona, 2015) for a very productive meeting. We are grateful to Anna Sippel, Anna Lisa Varri, and Paolo Bianchini for interesting discussions. We are grateful to Sverre Aarseth and Keigo Nitadori for making \textsc{nbody6} publicly available, and to Dan Foreman-Mackey for providing the \textsc{emcee} software and for maintaining the online documentation; we also thank Mr Dave Munro of the University of Surrey for hardware and software support. This research has made use of the products of the Cosmic-Lab project funded by the European Research Council. MG acknowledges financial support from the Royal Society (University Research Fellowship), AZ acknowledges financial support from the Royal Society (Newton International Fellowship). MP, MG and AZ acknowledge the European Research Council (ERC-StG-335936, CLUSTERS).

\bibliographystyle{mnras}

\label{lastpage}
\end{document}